# Temperature dependence of bend elastic constant in oblique helicoidal cholesterics

Olena S. Iadlovska[1,2], Greta Babakhanova[1,3], Georg H. Mehl[4], Christopher Welch[4], Ewan Cruickshank[5], Grant J. Strachan[5], John M. D. Storey[5], Corrie T. Imrie[5], Sergij V. Shiyanovskii[1,3], and Oleg D. Lavrentovich[1,2,3*]

[1]Advanced Materials and Liquid Crystal Institute, Kent State University, Kent, Ohio 44242, USA
[2]Department of Physics, Kent State University, Kent, Ohio 44242, USA
[3]Chemical Physics Interdisciplinary Program, Kent State University, Kent, Ohio 44242, USA
[4]Department of Chemistry, University of Hull, Hull HU6 7RX, United Kingdom
[5]Department of Chemistry, School of Natural and Computing Sciences, University of Aberdeen, AB24 3UE Scotland, United Kingdom
*e-mail address: olavrent@kent.edu

## Abstract

Elastic moduli of liquid crystals, known as Frank constants, are of quintessential importance for understanding fundamental properties of these materials and for the design of their applications. Although there are many methods to measure the Frank constants in the nematic phase, little is known about the elastic constants of the chiral version of the nematic, the so-called cholesteric liquid crystal, since the helicoidal structure of the cholesteric renders these methods inadequate. Here we present a technique to measure the bend modulus $K_{33}$ of cholesterics that is based on the electrically tunable reflection of light at an oblique helicoidal $Ch_{OH}$ cholesteric structure. $K_{33}$ is typically smaller than 0.6 pN, showing a non-monotonous temperature dependence with a slight increase near the transition to the twist-bend phase. $K_{33}$ depends strongly on the molecular composition. In particular, chiral mixtures that contain the flexible dimer 1″,7″-bis(4-cyanobiphenyl-4′-yl) heptane (CB7CB) and rod-like molecules such as pentylcyanobiphenyl (5CB) show a $K_{33}$ value that is 5 times smaller than $K_{33}$ of pure CB7CB or of mixtures of CB7CB with chiral dopants. Furthermore, $K_{33}$ in CB11CB doped with a chiral agent is noticeably smaller than $K_{33}$ in a similarly doped CB7CB which is explained by the longer flexible link in CB11CB.

The proposed technique allows a direct in-situ determination of how the molecular composition, molecular structure and molecular chirality affect the elastic properties of chiral liquid crystals.

## I. INTRODUCTION

Elastic constants of splay, twist and bend of liquid crystals define how these materials respond to external forces and boundary conditions [1,2]. There are many well-established methods of measuring elastic constants in the simplest type of liquid crystal, the so-called uniaxial nematic (N), see, for example, Refs. [3,4]. These methods typically use a monocrystalline sample in which the molecular orientation, specified by the director $\hat{\mathbf{n}}$ ($\hat{\mathbf{n}} \equiv -\hat{\mathbf{n}}, \hat{\mathbf{n}}^2 = 1$), is predesigned to be uniform in space. An external electric or magnetic field is applied to perturb this uniform orientation, and the elastic constants are deduced from the balance of the field strength and the elastic and surface anchoring forces that tend to preserve the initial alignment. These methods are hard to extend to the chiral type of the nematic phase, the cholesteric (Ch) phase, in which the director twists in space, remaining perpendicular to the helicoidal axis and thus forming a right-angle helicoid. The field response of this non-uniform ground state of the Ch phase involves complex structural reorganizations in which the director develops spatially varying twist and also deformations of splay and bend that are hard to separate from each other [1,2]. In absence of the direct measurements, it is usually assumed that the elastic properties of Ch phases are the same as those of their N counterparts. This assumption has never been tested. Thus, there is a clear need for a direct in-situ method to determine the elastic constants of Ch materials. Such a direct method is proposed in this work. It is applicable to Ch materials in which the external field creates a so-called oblique helicoidal structure (Ch$_{OH}$). By measuring the period of the structure as a function of the applied field, one deduces the bend modulus $K_{33}$.

The existence of the Ch$_{OH}$ state in a Ch acted upon by an electric or magnetic field has been envisaged theoretically a long time ago [5,6] and confirmed experimentally very recently [7-11]. The Ch$_{OH}$ structure occurs in chiral mixtures based on dimeric materials with a small bend constant $K_{33}$ as compared to the twist modulus $K_{22}$. Depending on the length of the methylene spacer, the smallest ratio $K_{33}/K_{22}$ for flexible dimers varies between 0.12 [12] and 0.16 [4]. In the presence of an electric $\mathbf{E}$ [7,8] or magnetic [9] field $\mathbf{H}$, the right-angle helicoid transforms into an oblique helicoid, where $\hat{\mathbf{n}}$ is tilted with respect to the helicoidal axis by some cone angle $\theta < 30°$ [7], thus forming the Ch$_{OH}$ state, Fig.1. The director $\hat{\mathbf{n}}$ in Ch$_{OH}$ experiences twist and bend deformations. The cone angle $\theta$ and the period $P$ of the structure decrease monotonously with the field increase, while preserving the single-harmonic periodic modulation of the director [7]. The pitch $P$ depends not only on $E$ but also on $K_{33}$ which allows us to use the dependence $P(E)$ for a direct measurement of $K_{33}$. The approach is illustrated for materials in which the application of an ac electric field $\mathbf{E}$ causes the Ch$_{OH}$ period $P$ to be in the submicron range; the value of $P$ is then easy to determine by studying selective Bragg reflection of light at the one-dimensional periodic structure of Ch$_{OH}$. The value of $K_{33}$ is deduced by measuring the optical response to the varying electric field. The method does not imply any extrapolation of the data from the non-chiral N state and represents a direct in-situ measurement of the bend elastic constant of a chiral liquid crystal.

The proposed method is tested for two types of cholesterics, one formed by a single-compound 1″,7″-bis(4-cyanobiphenyl-4′-yl) heptane (CB7CB) or 1″,11″- bis(4-cyanobiphenyl-4′-yl) undecane (CB11CB) with the flexible dimer molecules doped with a chiral dopant, and another representing a mixture containing a significant amount of rod-like molecules, pentylcyanobiphenyl (5CB), added to the flexible dimers. CB7CB doped with chiral additives yields a temperature behavior of $K_{33}$ in the Ch$_{OH}$ phase that is very close to the temperature behavior of $K_{33}$ in the N phase of pure CB7CB. Namely, $K_{33}$ decreases to low values of the order of 0.4 pN near the transition to the twist-bend phase in both N and Ch$_{OH}$. CB11CB doped with a chiral dopant shows an even smaller minimum value, $K_{33} = 0.2$ pN. Especially intriguing is the result that $K_{33}$ in the chiral mixtures that contain rod-like molecules of 5CB added to the flexible dimers is reduced to $K_{33} = 0.07$ pN.

## II. THEORETICAL MODEL

The peak wavelength $\lambda_{Bragg}$ of Bragg reflection of light at a uniform Ch$_{OH}$ structure is determined by the pitch $P$,

$$\lambda_{Bragg} = \bar{n}_{eff} P, \qquad (1)$$

and by the effective refractive indices

$$\bar{n}_{eff} = (n_o + n_{e,eff})/2 \quad \text{and} \quad n_{e,eff} = n_o n_e / \sqrt{n_e^2 \cos^2\theta + n_o^2 \sin^2\theta}, \qquad (2)$$

where $n_e$ and $n_o$ are the extraordinary and ordinary refractive indices, respectively. For small $\theta$, one can approximate $n_{e,eff} \approx n_o\left(1 + \frac{1}{2}\left(1 - \frac{n_o^2}{n_e^2}\right)\sin^2\theta\right)$ and

$$\bar{n}_{eff} = \frac{n_{e,eff} + n_o}{2} \approx n_o\left(1 + \frac{1}{4}\left(1 - \frac{n_o^2}{n_e^2}\right)\sin^2\theta\right). \qquad (3)$$

The pitch $P$ and the cone angle $\theta$ are both tunable by the applied electric field $E$ [7],

$$P = \frac{2\pi}{E}\sqrt{\frac{K_{33}}{\varepsilon_0 \Delta\varepsilon}}, \qquad (4)$$

$$\sin^2\theta = \frac{\kappa}{1-\kappa}\left(\frac{E_{NC}}{E} - 1\right); \quad \kappa = \frac{K_{33}}{K_{22}}, \qquad (5)$$

where $E_{NC} = 2\pi K_{22} / P_0\sqrt{\varepsilon_0 \Delta\varepsilon K_{33}}$ is a critical field at which Ch$_{OH}$ transforms into an unwound state, $\theta = 0$, $P_0$ is the equilibrium pitch of Ch in absence of the external field. Equation (4) suggests that $K_{33}$ can be determined from the dependence $P(E)$; the latter can be measured, for example, by using selective reflection of light.

Equations (4) and (5) have been derived assuming an ideal uniform Ch$_{OH}$. Such an ideal structure can exist in an infinitely thick sample, in which the surface anchoring of the director at the boundaries can be neglected. In cells of a finite thickness $d$, surface anchoring distorts the twist-bend director configuration and causes spatially-varying dielectric properties at the boundaries and redistribution of the electric field within the cell [10]. Selective reflection of light in these cells is determined by the central bulk region, in which $P$, $\theta$, and the acting electric field $E_{bulk}$ are coordinate-independent [10]. Thus, in Eqs. (4) and (5), the variable $E$ should be replaced with $E_{bulk}$.

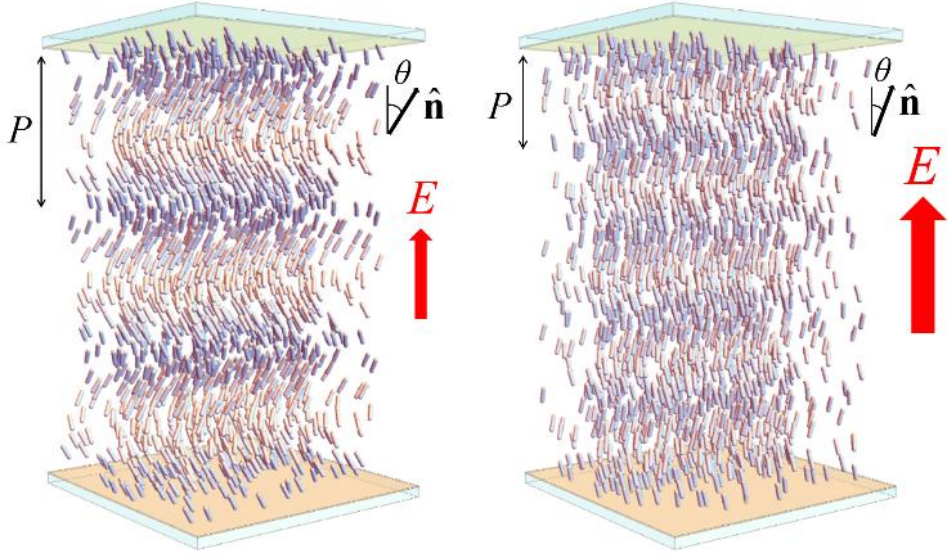

FIG. 1. The oblique helicoidal structure with the pitch $P$ and the cone angle $\theta$ both tunable by the applied electric field **E**.

As demonstrated in Ref. [10], the subsurface inhomogeneous regions make the acting electric field $E_{bulk}$ different from the average applied electric field $E_{av} = U/d$, where $U$ is the applied voltage:

$$E_{bulk} = E_{av} / (1+\xi) \;;  \qquad (6)$$

$\xi = \alpha E_{av}^{-1}$ is a small correction factor related to the dielectric extrapolation length and $\alpha$ is the adjusting coefficient. However, as we shall see later, corrections associated with $\xi$ are negligibly small, yielding only 1% uncertainty of the deduced values of $K_{33}$.

Using Eqs. (1) and (2) and accounting for the finite cell thickness (6), the wavelength of the reflection peak $\lambda_{Bragg}$ for the known applied electric field $E_{av}$ can be written as

$$\frac{\lambda_{Bragg}}{n_o} = a_1 E_{av}^{-1} + a_2 E_{av}^{-2} + O\left(E_{av}^{-3}\right), \tag{7}$$

where $a_1 = 2\pi \sqrt{\frac{K_{33}}{\varepsilon_0 \Delta \varepsilon}} \left(1 - \frac{\kappa}{4(1-\kappa)}\left(1 - \frac{n_o^2}{n_e^2}\right)\right)$ and $a_2 = \alpha a_1 + \sqrt{\frac{K_{33}}{\varepsilon_0 \Delta \varepsilon}} \frac{\kappa \pi E_{NC}}{2(1-\kappa)}\left(1 - \frac{n_o^2}{n_e^2}\right)$.

### III. EXPERIMENTAL METHODS AND MATERIALS

We studied four chiral mixtures. All contain flexible dimeric molecules that are known to form bend conformations and as a result, induce the so-called twist-bend nematic phase with nanoscale director modulation [13-15], the Ch$_{OH}$ state [7,8] and a small $K_{33}$ in the nematic phase [4,12,16]. All these mixtures feature a Ch phase and a chiral analog of the twist-bend nematic phase ($N_{TB}^*$) [13,14,17].

The first mixture, abbreviated CB7CB:S811, represents a well-studied flexible dimer 1″,7″-bis(4-cyanobiphenyl-4′-yl) heptane (CB7CB) for which the elastic constants have been determined previously [4], doped with a left-handed chiral additive S811 (EM Industries) in weight proportion CB7CB:S811 = 96:4. The cholesteric phase of CB7CB:S811 shows a transition into an isotropic fluid at $T_{Ch-I} = 110.8°C$ and into $N_{TB}^*$ at $T_{TB} = 95.5°C$, Fig.2. The temperature was controlled by a hot stage HCS402 and a controller mK2000 (both Instec, Inc.) with the accuracy of $0.1\,°C$.

The second mixture CB11CB:S811 is formulated using a longer methylene-linked dimer 1″,11″-bis(4-cyanobiphenyl-4′-yl) undecane (CB11CB) [18] doped with S811 in weight proportion CB11CB:S811 = 97:3. This mixture shows the transition temperatures $T_{Ch-I} = 125.3°C$ and $T_{TB} = 107.3°C$.

The remaining two studied chiral materials are based on a nematic mixture M0 that contains flexible dimers CB7CB, CB11CB and a rod-like mesogen 5CB (EM Industries) in weight proportion 52:31:17. The phase diagram of M0 is I (64 °C) N (28.3 °C) $N_{TB}$; here and elsewhere the phase diagrams and other experimental data were obtained in the regime of cooling. This mixture was doped with two different amounts of S811, 1.8wt% (mixture M1.8) and 4.2% (mixture M4.2). The phase diagram of M1.8 is I (61.3 °C) Ch (27.3 °C) $N_{TB}^*$, while for M4.2, it is I (60.7 °C) Ch (24.7 °C) $N_{TB}^*$.

We used planar cells assembled from two glass plates coated with transparent indium tin oxide (ITO) electrodes. The cells of a gap thickness $d$ = (16.0-24.0) μm were purchased from EHC Co, Japan, or prepared in the Advanced Materials and Liquid Crystal Institute laboratory; in the latter case, planar alignment was achieved by a rubbed layer of polyimide PI2555 (Nissan Chemicals, Ltd.). The cell thickness was set by spherical spacers mixed with UV-curable glue NOA 68 (Norland Products, Inc.) and measured with a UV/VIS spectrometer (Perkin Elmer, Inc).

All studied materials show very good planar alignment. Figure 3 shows a field-free Ch planar texture of a cell with the mixture M4.2 used for measurements of dielectric permittivities, while Figure 4 shows the textures of the Ch$_{OH}$ state for the same mixture used for spectral measurements. The planar alignment is not deteriorated by the temperature changes within the range of stability of the Ch phase. The only exception is the narrow region, less than 1 °C, near the Ch-$N_{TB}^*$ transition of the CB7CB:S811 mixtures, in which the uniform planar Ch texture experiences undulations associated with the decrease of the cholesteric pitch $P_0$, Fig. 2(a).

The sinusoidal ac field of frequency 3 kHz was applied using a DS345 waveform generator (Stanford Research) and 7602-M wideband amplifier (KROHN-HITE Co.). The electric field **E** induces the oblique helicoidal Ch$_{OH}$ structure with its axis $\hat{\mathbf{t}}$ along the field, $\hat{\mathbf{t}} \parallel \mathbf{E}$. At a fixed temperature, the Ch$_{OH}$ pitch $P$ is tunable by the field in a wide spectral range including the visible part, Fig.7(a) [7,8].

The Bragg reflection was recorded using a tungsten halogen light source LS-1 and USB2000 spectrometer (both Ocean Optics). The light source generates an unpolarized beam focused by the lens into a paraxial ray incident normally on the surface of the cell. The wavelength $\lambda_{Bragg}$ of the reflection peak was determined by measuring the bandwidth of the peak $\Delta\lambda$ at its half-amplitude, and then defining $\lambda_{Bragg}$ as the coordinate of the middle of the bandwidth. The refractive indices of the nematic CB11CB and the mixture M0 were measured using the wedge cell technique [19].

The dielectric characterization of the materials was performed using an LCR meter 4284A (Hewlett Packard) and the GenRad 1628 Capacitance Bridge (IET Labs). The perpendicular permittivity $\varepsilon_\perp$ of the nematic CB11CB and Ch mixtures M1.8 and M4.2 was determined by measuring capacitance in planar cells, Fig. 3, at low applied voltage that does not perturb the planar structure. The parallel permittivity $\varepsilon_\parallel$ was determined in the same cells, by applying a high voltage to unwind the Ch into the field-aligned state, and then calculating $\varepsilon_\parallel$ by the extrapolation method described elsewhere for the N phase [20].

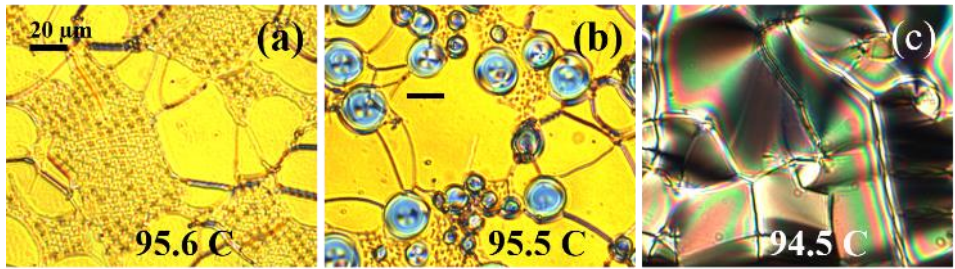

FIG. 2. Phase transition of CB7CB:S811 upon cooling, from (a) the Ch phase with undulations at 95.6 °C into (b) biphasic Ch - $N_{TB}^*$ state with nucleating $N_{TB}^*$ islands at $T_{TB} = 95.5$ °C; (c) complete transformation into $N_{TB}^*$ at 94.5 °C.

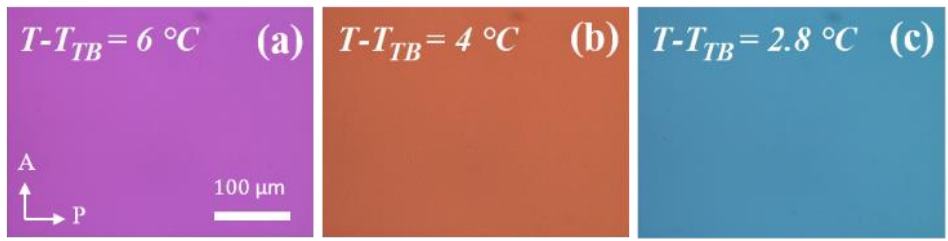

FIG. 3. Uniform planar Ch texture of mixture M4.2 in the absence of field at temperatures (a) $T - T_{TB} = 6$ °C, (b) $T - T_{TB} = 4$ °C, and (c) $T - T_{TB} = 2.8$ °C observed in transmission mode of the polarizing optical microscope (POM) with crossed polarizer (P) and analyzer (A) on cooling.

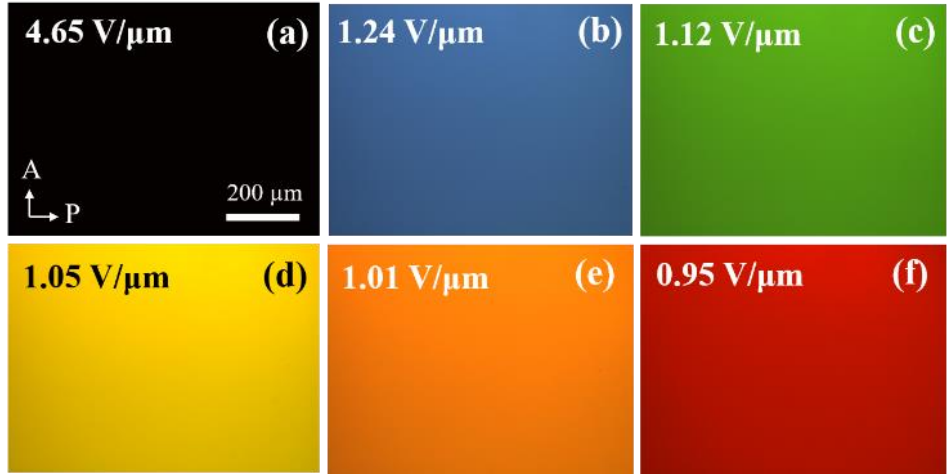

FIG. 4. Polarizing microscopy textures of mixture M4.2 in planar cell subject to an applied electric field, at a fixed temperature $T - T_{TB} = 2.8\ °C$: (a) the field-aligned N state observed in transmission; (b) – (f) uniform Ch$_{OH}$ textures with field-controlled structural colors observed in the reflection mode. The field RMS amplitude is indicated in the figures.

## IV. RESULTS AND DISCUSSION

### A. Optical and dielectric properties of the materials

*CB7CB*. Optical and dielectric properties of the dimer CB7CB have been reported previously [4]. In order to determine the linear coefficient $a_1$ in Eq. (7), we measured the temperature dependence of the ordinary refractive index $n_o$ at 450 nm, 532 nm and 633 nm in the range $0 \leq T - T_{TB} \leq 6°C$. Within this range, $n_o$ is practically temperature-independent, with $n_o = 1.578$ at 450 nm, $n_o = 1.562$ at 532 nm and $n_o = 1.554$ at 633 nm. These values are used to determine the dispersion through the Cauchy formula,

$$n_o(\lambda) = A + B\lambda^{-2} + C\lambda^{-4}, \tag{8}$$

where the polynomial coefficients are determined by fitting the experimental data, $A = 1.549$, $B = -2.05 \cdot 10^{-3}\ \mu m^2$ and $C = 1.60 \cdot 10^{-3}\ \mu m^4$.

***CB11CB.*** We measured the temperature dependencies of $n_o$ and $n_e$ at $\lambda = 532$ nm, and also the temperature dependence of $n_o$ at 450 nm and 633 nm, Fig. 5(a). In the range $0 \leq T - T_{TB} \leq 6°C$, $n_o$ is practically temperature-independent, $n_o = 1.550$ at 450 nm, $n_o = 1.533$ at 532 nm, and $n_o = 1.524$ at 633 nm. The data are extrapolated by Eq. (8) with $A = 1.510$, $B = 2.82 \cdot 10^{-3}$ μm$^2$, and $C = 1.05 \cdot 10^{-3}$ μm$^4$. The measured temperature dependencies of $\varepsilon_\parallel$, $\varepsilon_\perp$ and $\Delta \varepsilon = \varepsilon_\parallel - \varepsilon_\perp$ for the nematic CB11CB are shown in Fig. 5(b).

***M0.*** The measured $n_o, n_e$ in the nematic M0 are presented in Fig. 6(a). In the range $0 \leq T - T_{TB} \leq 6°C$, $n_o$ is again temperature-independent, $n_o = 1.570$ at 450 nm, $n_o = 1.548$ at 532 nm, and $n_o = 1.538$ at 633 nm, which allows the data to be extrapolated by the Cauchy expansion, Eq. (8), with $A = 1.532$, $B = -2.91 \cdot 10^{-3}$ μm$^2$, and $C = 2.14 \cdot 10^{-3}$ μm$^4$. The temperature dependence of the dielectric anisotropy $\Delta \varepsilon(T)$ was measured separately in the nematic M0 and the chiral mixtures M1.8, M4.2, Fig. 6(b).

Note that near the phase transition to the twist-bend nematic phase $N_{TB}$, both CB11CB and M0 show a non-monotonous change of $n_e$, with a shallow minimum, Fig. 5(a) and 6(a), respectively. Similar behavior is noted for the birefringence in other dimeric nematics [13,21].

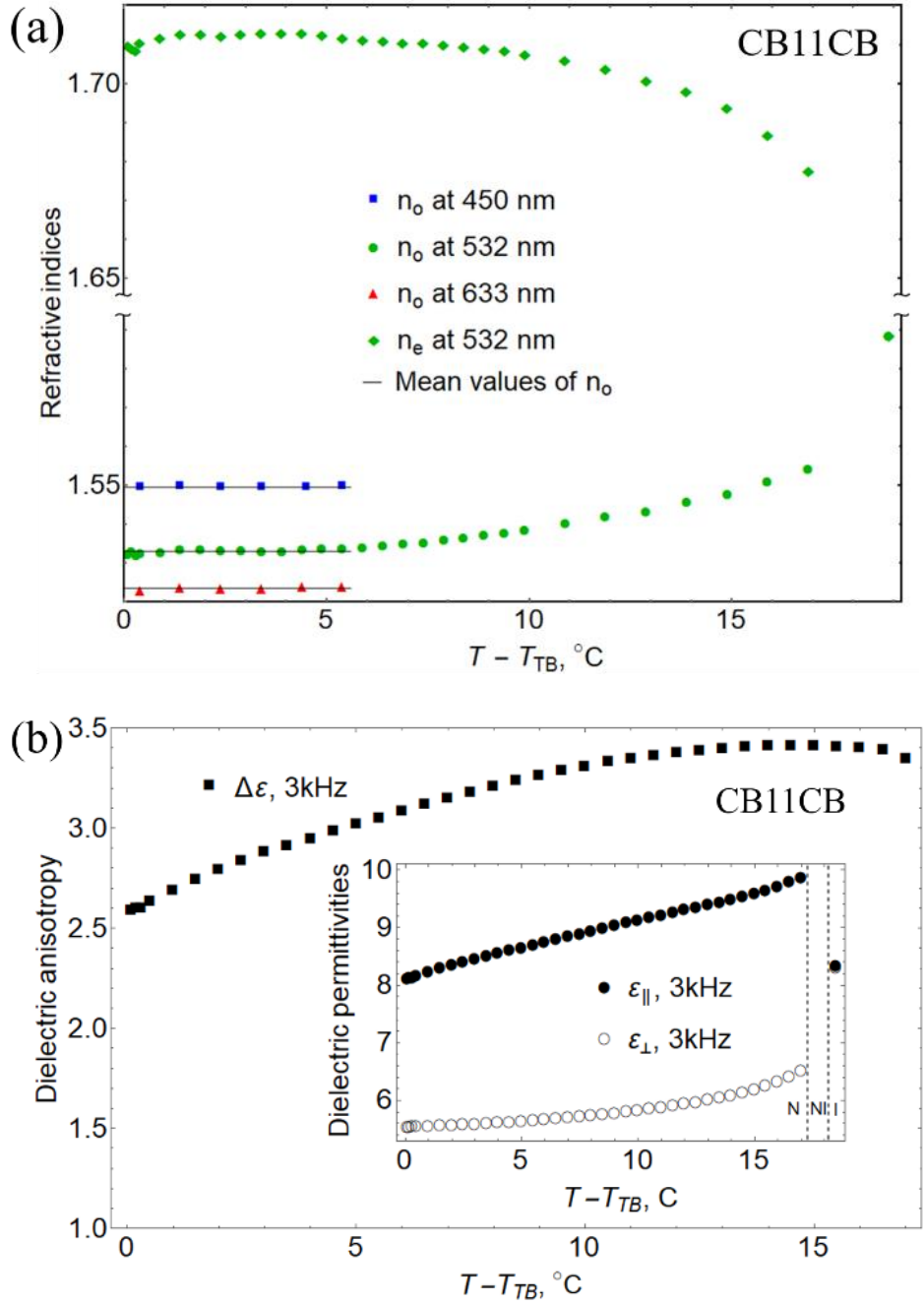

FIG.5. Temperature dependencies of the material parameters of CB11CB: (a) extraordinary $n_e$ refractive index at 532 nm and the ordinary refractive index $n_o$ at 450 nm, 532 nm, and 633 nm; (b) dielectric permittivities and dielectric anisotropy.

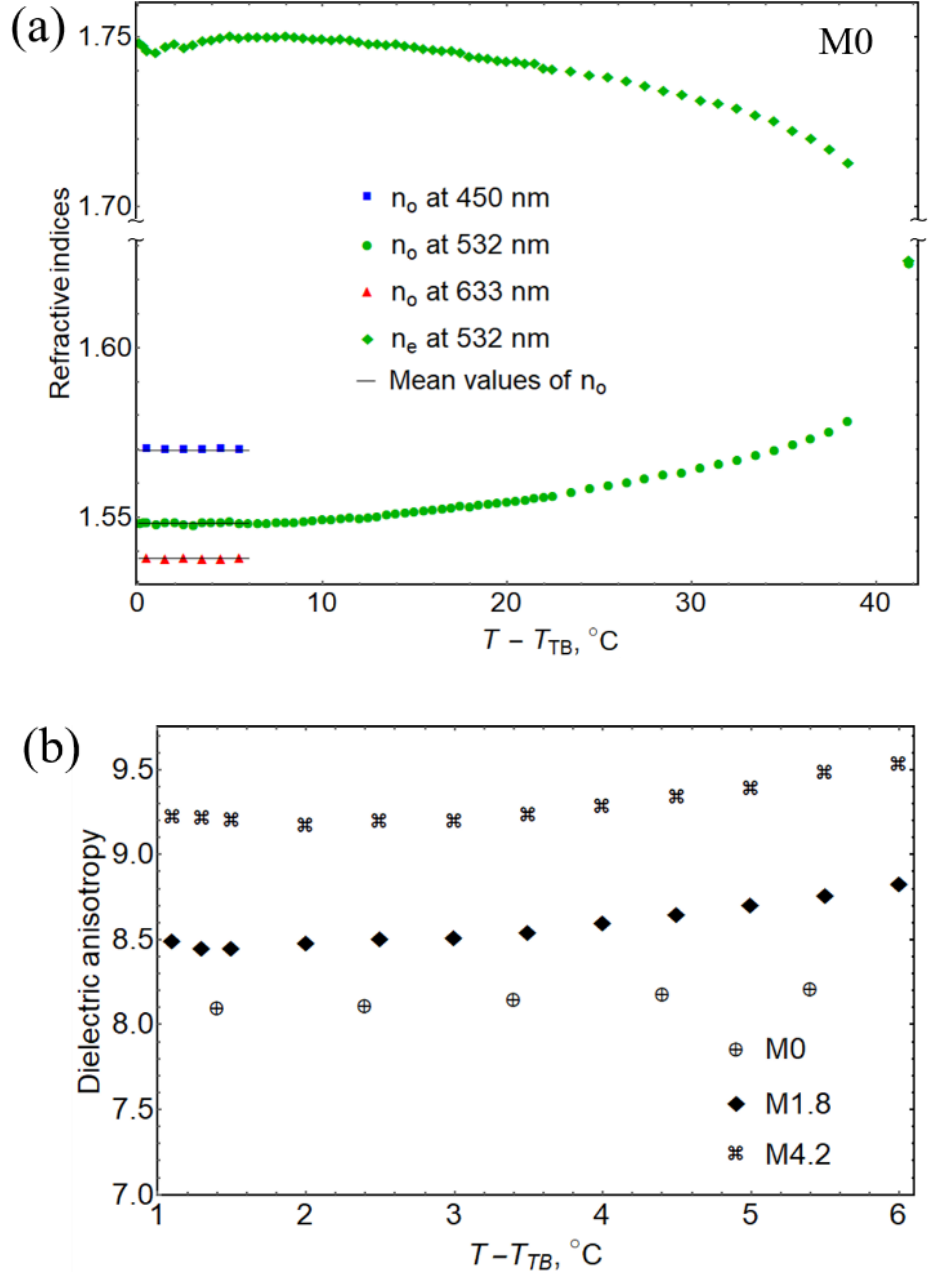

FIG. 6. Temperature dependencies of the material parameters of nematic M0 and cholesterics M1.2, M4.2: (a) $n_e$ at 532 nm and $n_o$ at 450 nm, 532 nm, and 633 nm of M0; (b) the dielectric anisotropy $\Delta\varepsilon(T)$ measured in three mixtures, nematic M0 (⊕), and cholesteric M1.8 (♦) and M4.2 (⌘).

## B. Bragg reflection spectra of Ch$_{OH}$

The Ch$_{OH}$ state forms under an applied electric field at temperatures above the $N^*_{TB}$-Ch phase transition. The Bragg reflection spectra are measured as a function of the electric field $E_{av} = U/d$. The reflection peaks were collected from well equilibrated (over 10 min) Ch$_{OH}$ states. An example of the field-controlled spectral properties of M4.2 is shown in Fig.7(a). These spectra are used to determine the wavelength $\lambda_{Bragg}$ of Bragg reflection at each applied field $E_{av}$.

The Bragg reflection peak shape depends on the wavelength and applied electric field, since the cone angle changes with the field, Fig.1 and Eq. (5). For example, at high fields, the cone angle is small, thus the peaks in the blue part of the spectrum are narrow, while at low fields, the cone angle is larger and the peaks are correspondingly wider and flatter in the red part of the spectrum. Because of the complicated interplay between the peak shape and the material properties, we calculate the Bragg wavelength $\lambda_{Bragg}$ as the middle coordinate of the peak's width at its half-amplitude. The accuracy of the $\lambda_{Bragg}$ measurement is estimated to be better than 0.5 nm. After $\lambda_{Bragg}$ and the corresponding $E_{av}$ are determined, the quantity $\lambda_{Bragg}/n_o(\lambda_{Bragg})$ is plotted vs. $E_{av}^{-1}$ for each temperature point and fitted with the polynomial $a_1 E_{av}^{-1} + a_2 E_{av}^{-2}$, see Eq. (7) and Fig. 7(b). As clearly evidenced by the linear fit in Fig.7(b), the linear coefficient $a_1$ is much larger than the quadratic contribution $a_2 E_{av}^{-1}$, namely, $a_1/a_2 E_{av} \sim 10^2$, which means that the finite cell thickness correction (6) has practically no influence on the $K_{33}$ values. Thus, Eq. (7) can be significantly simplified by dropping the $a_2$ term:

$$K_{33} = \frac{\varepsilon_0 \Delta\varepsilon a_1^2}{4\pi^2 \left(1 - \frac{\kappa}{4(1-\kappa)}\left(1 - \frac{n_o^2}{n_e^2}\right)\right)^2}. \qquad (9)$$

In Eq. (9), $K_{33}$ enters both sides of the equation, since $\kappa = K_{33}/K_{22}$. Thus, it is important to establish whether a knowledge of the twist constant $K_{22}$ is necessary to determine $K_{33}$, which amounts to the question of how close the factor $Z = \left(1 - \frac{\kappa}{4(1-\kappa)}\left(1 - \frac{n_o^2}{n_e^2}\right)\right)^2$ is to 1. As demonstrated below, $Z$ differs very little from 1, typically by less than 2%.

The twist elastic constant in the nematic phase of CB7CB measured within about $6\,^\circ\mathrm{C}$ from the transition to the twist-bend nematic phase, is $K_{22} = (2.5 \div 3)\,\mathrm{pN}$, while the maximum value of the bend constant is $K_{33} = 0.5\,\mathrm{pN}$ [4]. As a result, the highest value of $\kappa$ is 0.2 and $\frac{\kappa}{4(1-\kappa)} \approx 0.063$. Furthermore, the measured indices of refraction for CB7CB (at 633 nm) in the same temperature range are $n_o = 1.56$ and $n_e = 1.69$, thus $1 - n_o^2/n_e^2 = 0.15$. Therefore, $Z = \left(1 - \frac{\kappa}{4(1-\kappa)}\left(1 - \frac{n_o^2}{n_e^2}\right)\right)^2 \approx 0.98$ is very close to 1. In other words, the uncertainty in the measurements of $K_{33}$ caused by the presence of $K_{33}$ in the right-hand side of Eq. (9) is less than 2% for CB7CB:S811. Thus, for the low-temperature materials M1.8 and M4.2 the error is even smaller, less than 1%, since the numerical solution of Eq. (9) leads to the values of $K_{33}$ of the order of 0.1 pN, five times smaller than in CB7CB:S811. In what follows, we assume $Z = 1$ for all materials and use the following simple expression to determine $K_{33}$ experimentally:

$$K_{33} = \frac{\varepsilon_0 \Delta\varepsilon}{4\pi^2} a_1^2. \qquad (10)$$

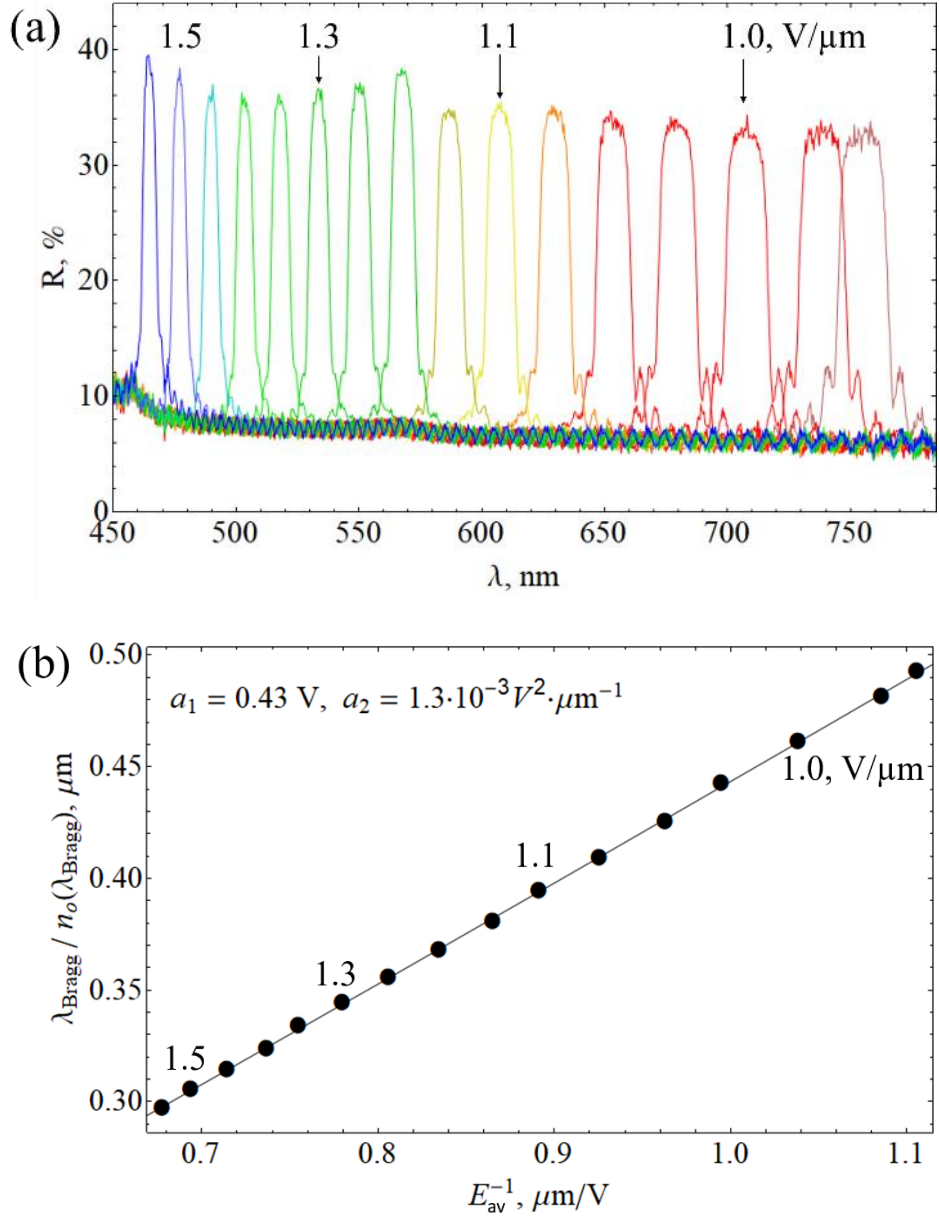

FIG. 7. (a) Bragg reflection spectra and (b) the corresponding dependence of the normalized Bragg wavelengths $\lambda_{Bragg}/n_o(\lambda_{Bragg})$ vs the inverse electric field for the Ch$_{OH}$ mixture M4.2 measured at $T-T_{TB}=6\ °C$. The experimental points $\lambda_{Bragg}/n_o(\lambda_{Bragg})$ are fitted with the polynomial $a_1 E_{av}^{-1}+a_2 E_{av}^{-2}$, Eq. (7); the error bars are smaller than the symbols size. In part (a), some reflection peaks are supplemented with the corresponding electric field values; these values are also presented in part (b).

### C. Bend elastic constant of Ch$_{OH}$

According to Eq. (10), $K_{33}$ is determined by the fitting parameter $a_1$ of the electric field dependence of the Bragg wavelength $\lambda_{Bragg}$ and by the dielectric anisotropy $\Delta\varepsilon$. Using the data collected for all four chiral mixtures, we determined the temperature dependencies of $K_{33}$, Fig. 8, in the temperature range $0 \leq T - T_{TB} \leq 6°C$. The $K_{33}$ modulus of M1.8 and M4.2 cannot be measured in the range $0 \leq T - T_{TB} \leq 1°C$, because at low temperatures, the selectively reflective textures of these mixtures do not equilibrate to a single-peak state after many hours or relaxation [22].

The temperature dependence of $K_{33}$ in the chiral CB7CB:S811 mixture is very similar to that previously determined by Babakhanova *et al* [4] for a pure nematic CB7CB and also for other flexible dimers [12,16]. Namely, $K_{33}$ in CB7CB:S811 decreases as one moves towards the transition into the twist-bend state, down to about 0.4 pN; while at $T \leq T_{TB} + 1°C$, $K_{33}$ shows a pretransitional increase. In CB7CB:S811, the dependence $K_{33}(T)$ is slightly shifted down along the temperature axis in the pretransitional region, as compared to a nematic CB7CB, Fig. 8. The pretransitional increase in $K_{33}(T)$ in materials such as CB7CB is usually associated with the appearance of pretransitional molecular clusters of the twist-bend phase; in these clusters, equidistance of one-dimensional director modulations hinders deformations of twist and bend of the twist-bend axis [13,16,23]. The chiral dopant might suppress the development of these clusters, thus reducing the pretransitional increase of $K_{33}$ in CB7CB:S811 as compared to that of pure CB7CB in Fig. 8.

The chiral mixture CB11CB:S811 near the transition to $N_{TB}^*$ exhibits $K_{33}$ that is noticeably smaller than $K_{33}$ in CB7CB:S811. The minimum value is $K_{33} = 0.16 \text{ pN}$, which is 2.4 times smaller than the minimum value in CB7CB:S811. This difference can be attributed to the longer methylene bridge of CB11CB dimers, which might contribute to their higher bend flexibility as compared to CB7CB molecules.

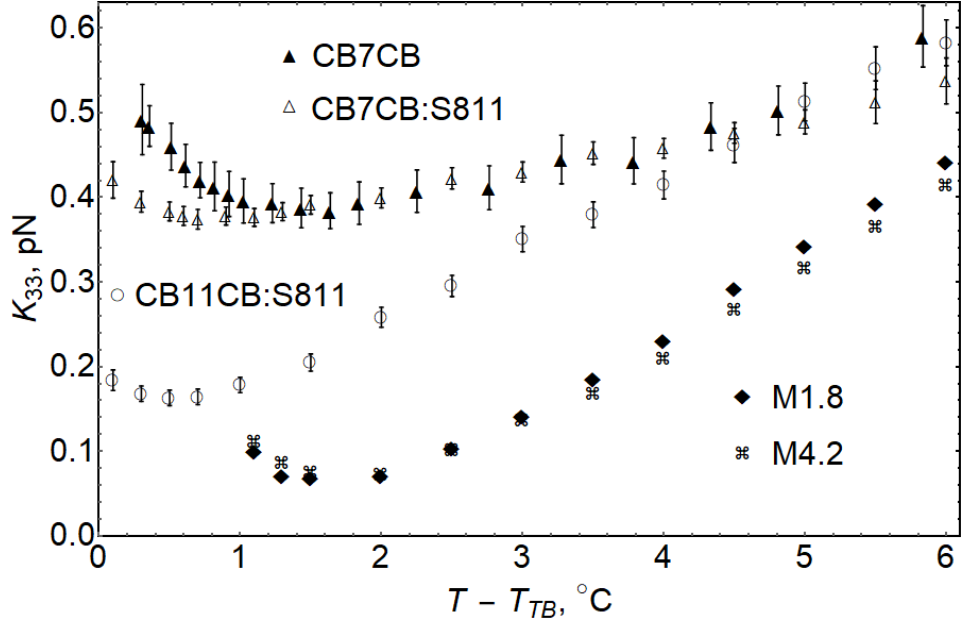

FIG. 8. The temperature dependencies of the bend elastic constant $K_{33}(T)$ for nematic CB7CB (▲) (replotted from Ref. [4]), and chiral mixtures CB7CB:S811 (△), CB11CB:S811 (○), M1.8 (♦) and M4.2 (⌘).

In the mixtures M1.8 and M4.2, $K_{33}(T)$ shows a similar non-monotonous behavior. A rather unexpected feature is that in these mixtures that contain 5CB rod-like molecules, the minimum value of $K_{33}$ is extremely low, about 0.07 pN, i.e., 5.4 times smaller than in CB7CB:S811 and 2.3 times smaller than in CB11CB:S811. The exact mechanism accounting for this difference is not clear. Tentatively, it may be associated with the geometry of the 5CB molecules that represent approximately one half of the CB7CB and CB11CB molecules. In the mixtures, 5CB molecules might serve as broken CB7CB/CB11CB dimers that facilitate the bend of the latter. Obviously, this argument cannot be extended to the pure nematic phase of 5CB, in which $K_{33} \sim 10$ pN [24]. Finally, it is of interest to note that at $T > T_{TB} + 3°C$, $K_{33}$ in a strongly chiral mixture M4.2 is slightly smaller than in the weakly chiral M1.8.

## V. CONCLUSION

We propose a direct method of measuring the bend elastic modulus $K_{33}$ in chiral nematics. $K_{33}$ is calculated from the position of Bragg peaks $\lambda_{Bragg}$ for equilibrium $Ch_{OH}$ states, which are well-determined at fixed values of applied field. The method is applicable to the range of temperatures and the electric fields at which the $Ch_{OH}$ structure is stable. According to the theory proposed by Xiang *et al* [7], the upper $E_{NC}$ and lower $E_{N*C}$ fields of the stable $Ch_{OH}$ structure are related, here $E_{N*C} \approx \kappa E_{NC}\left[2+\sqrt{2(1-\kappa)}\right]/(1+\kappa)$ is the field at which the $Ch_{OH}$ state with the helical axis $\hat{\mathbf{t}}$ parallel to the field $\mathbf{E}$ transforms into a right-angle helicoidal Ch state with the helical axis $\hat{\mathbf{t}}$ perpendicular to $\mathbf{E}$ [7]. To keep $E_{NC} > E_{N*C}$, the ratio $\kappa = K_{33}/K_{22}$ must be lower than 0.5. Obviously, the method is applicable only to the materials with positive dielectric anisotropy, since $Ch_{OH}$ does not form in the materials with negative dielectric anisotropy. In principle, the method based on $Ch_{OH}$ structures can be expanded to the measurements of the twist modulus $K_{22}$, as follows from Eq. (4) and the definition of the critical field $E_{NC}$, if $E_{NC}$ and the Ch pitch $P_0$ are measured independently. In the studied mixtures, the fields close to $E_{NC}$ shift the optical response too far into the UV region to make measurements by spectral studies accurate. We leave the goal of simultaneous measurements of both $K_{33}$ and $K_{22}$ to future studies.

The bend constant $K_{33}$ shows a nonmonotonous dependence on the temperature near the transition to the chiral analog of the twist-bend nematic phase, with a pronounced minimum and an increase as the temperature is lowered. This feature is universal for both nematic and chiral nematics that show a transition into the twist-bend state and can be associated with the appearance of pretransitional molecular clusters of the twist-bend phase; in these clusters, equidistance of one-dimensional director modulations hinders deformations of twist and bend of the twist-bend axis [13,16,23]. The dimer CB7CB doped with the chiral additive S811 exhibits a temperature behavior of $K_{33}$ that is very close to that in the nematic phase of pure CB7CB. The longer methylene-linked CB11CB dimer doped with S811 shows $K_{33}$ with the minimum value 2.4 times smaller than its counterpart CB7CB:S811, which might be related to a lower resistance to bending of the long methylene bridge in the CB11CB molecule. Our data on the bend constant $K_{33}$ in the mixture

CB11CB:S811 are significantly lower than $K_{33}$ measured previously by Balachandran *et al* [18] for pure CB11CB. In our case, $K_{33}$ is in the range (0.16 ÷ 0.55) pN, depending on temperature, Fig. 8. In contrast, Ref. [18] reports $K_{33}$ in the range (5.8 ÷ 10.5) pN, at least one order of magnitude higher. Note here that CB11CB shows the $N_{TB}$ phase, the existence of which requires a very small value of $K_{33}$. In fact, the first measurements of the elastic constants in $N_{TB}$-forming flexible dimeric materials [3,4,12,13,16] show that $K_{33}$ is below 1 pN within a range of about 5 – 10 °C above the N-$N_{TB}$ phase transition. Our data for the CB11CB:S811 mixtures are in line with these previous measurements of $K_{33}$ in flexible dimeric materials.

Rather surprisingly, we also observe a dramatic decrease of $K_{33}$ in the chiral mixtures that contain rod-like 5CB molecules added to flexible dimers; in these mixtures, the minimum value is $K_{33} = 0.07$ pN, which is 5.4 times smaller than the minimum value in CB7CB:S811. A tentative explanation is that the relatively short 5CB molecules serve as structural "fillers" that facilitate the bend of CB7CB/CB11CB dimers. The detailed mechanisms of how molecular composition, molecular structure and chirality influence the observed features of macroscopic elastic properties of the chiral nematic phases are not known and deserve further studies.

## Acknowledgements

The work was supported by the NSF grant ECCS-1906104. The authors would like to thank Peter Palffy-Muhoray and Tianyi Guo for kindly providing the equipment for capacitance measurements at high fields. We are thankful to the Reviewers for useful suggestions. The work of G.H.M and C.W. was supported by the EPSRC grant EP/M015726.